\documentstyle[12pt]{article}

\begin{document}

% The general mathematics macros
% ************************************* MACROS

% bold mu
\def\b #1{\mbox{\footnotesize\boldmath$#1$\normalsize}}

% bold mu barra
\def\bb#1{\mbox{\footnotesize\boldmath$\overline{#1}$\normalsize}}

% Gran delta
\def\DG{\mbox{\large$\delta$\normalsize}}

% ************************************* 0
\title{\hspace{10cm}
{\footnotesize{IFFC preprint 96-01}}\\
Extended knots and the space of states of quantum
gravity\footnote{submitted to Nucl. Phys. B}}

\author{\small Jorge Griego \\
\small Instituto de F\'{\i}sica, Facultad de Ciencias,\\
\small Trist\'an Narvaja 1674, 11200 Montevideo, Uruguay.\\
\small E-mail: griego@fisica.edu.uy  \\ \\}

\date{January 6, 1996}
\maketitle
\abstract{In the loop representation the quantum constraints of gravity
can be solved. This fact allowed significant progress in the
understanding of the space of states of the theory.  The analysis of the
constraints over loop dependent wavefunctions has been traditionally
based upon geometric (in contrast to analytic) properties of the loops.
The reason for this preferred way is twofold: for one hand the inherent
difficulties associated with the analytic loop calculus, and on the
other our limited knowledge about the analytic properties of knots
invariants. Extended loops provide a way to overcome the difficulties at
both levels. For one hand, a systematic method to construct analytic
expressions of diffeomorphism invariants (the extended knots) in terms
of the Chern-Simons propagators can be developed. Extended knots are
simply related to ordinary knots (at least formally). The analytic
expressions of knot invariants could be produced then in a generic way.
On the other hand, the evaluation of the Hamiltonian over extended loop
wavefunctions can be thoroughly accomplished in the extended loop
framework. These two ingredients promote extended loops as a potential
resort for answering important questions about quantum gravity.}

\section{Introduction}
Knot theory and quantum gravity have a profound relationship that is put
into manifest within the framework of the Ashtekar new variable
reformulation of general relativity \cite{As}. The facts that {\it i)} a
loop representation of the Lie-algebra valued Ashtekar connection can be
introduced \cite{RoSm}; {\it ii)} the quantum states of gravity are
given in terms of knot invariants due to the diffeomorphism invariance
of the theory; {\it iii)} there exist knot invariants that are solutions
of the Wheeler-DeWitt equation \cite{RoSm,BrGaPu}; clearly show the
capability of knots to capture relevant information about the quantum
properties of space-time.

The relationship between knots and physics goes far beyond the interest
of general relativity \cite{Kabook}. In fact, the many ways that
knot theory interacts with physics has motivated a renewed interest of
physicists to adventure into the traditionally mathematician world of
knots (the first goal was in the physics of the late 1800's with the
theory of vortex atoms of Lord Kelvin). Part of the recent work was
devoted to study in depth the significant relationship between
topological quantum field theories and knot polynomials
\cite{Wi,GuMaMi,BF}, and in particular for the quantum gravity case the
specific relation between some knot polynomials and the solutions of the
Hamiltonian constraint in the loop representation \cite{GaPubook}.

In gravitation, until the appearance of the results of Br\"ugmann,
Gambini and Pullin \cite{BrGaPu2} the calculations with loop
dependent objects were mainly supported on geometric (rather than
analytic) basis. For example, the geometric properties of the smoothened
loops (i.e. the absence of kinks and intersections) underlies the
topological construct (weaves) that bridges the usual metric tensor with
the discrete properties of spacetime at the Planck length \cite{AsRoSm}.
The solution obtained by Br\"ugmann et al. was the first analytic (in the
sense that it does not depend of any particular geometric property of
the loops) and also the first to correspond to a nondegenerate state. Up
today no others solutions of this type have been found.

The loop calculus faces up to significant difficulties at the
analytic level that, for quantum gravity, have two features. For one
hand, the knowledge of the analytic properties of the ``kinematical"
space of states (the knots) is very limited. On the other, any
explicit analytic evaluation of the Hamiltonian constraint over
diffeomorphism invariant loop wavefunctions involves hard technical and
conceptual problems. The technical difficulties refer to both formal and
regulated calculations. The conceptual ones concern the interpretation
of a derivative operator over topological objects (see \cite{DiGaGrPu2}
for a discussion of this last point). This means that, in spite of the
loop representation has had the virtue to unlock the canonical
quantization program of gravity, the identification of the space of
states of the theory in terms of functional of loops is at present far to
be complete.

Extended loops offer a new avenue to this problem. The extended loops
were introduced as suitable generalizations of ordinary loops (the usual
group of loops is embedded into a local infinite dimensional Lie group,
the extended loop group \cite{DiGaGr1}). In resemblance with ordinary
loops, an extended loop representation of any Lie-algebra valued
connection theory can be constructed. In particular, the extended loop
representation of the Ashtekar connection was recently developed
\cite{DiGaGrPu1,DiGaGr2}. This representation contains (and it is simply
related to) the conventional loop representation of quantum gravity. As
the extended loop group has a more rich mathematical structure than the
usual group of loops, some benefits at the calculation and
regularization levels are exhibited by the new representation. These
advantages has been proved to be relevant in the study of the
Wheeler-DeWitt equation \cite{Gr5,Gr6,Grprep}.

The aim of this article is to develop a systematic method to obtain
analytic expressions of diffeomorphism invariants (the extended knots).
The general ideas of the method were advanced in \cite{Gr9}. Now we are
going to study with some detail the mechanism of operation of the
diffeomorphism constraint in terms of extended loops, and we will
establish an explicit procedure to get solutions. Extended knots reduce
to ordinary knots by means of a simple prescription (at least at the
formal level). The application and potential significance of the method
to the case of quantum gravity are discussed.

The article is organized as follows: in Sect. 2 the properties of the
extended loop wavefunctions are considered from a general point of view.
In Sect. 3 the rudiments of a method to construct analytic expressions
of diffeomorphism invariants in terms of extended loops are introduced.
The existence of a systematic for the operation of the constraints in
the extended loop framework was first advanced in \cite{Gr6}. In Sect. 4
this systematic is applied to build up {\it families} of extended knots
invariants. We define a family as a set of diffeomorphism invariants
with a specific structure. In Sect. 5 we consider a particular family.
The analytic expressions of the members of this family are derived and
their identification with known knot invariants is performed. The
Mandelstam identities of the members of the family are also analyzed.
The implications of the results for quantum gravity are discussed in
Sect. 7. This Section also includes the conclusions. An appendix with
some useful results is added.

\section{Extended loop wavefunctions}
Ordinary loops can be codified by an infinite set of distributional
fields, the multitangent fields. The multitangent fields are defined by
the integration of an ordered sequence of distributions along a loop
$\gamma$ in the following way

\begin{equation}
X^{\mu_1 \ldots \mu_r}(\gamma) :=
\oint_\gamma dy_r^{a_r} \! \ldots \! \oint_\gamma dy_{1}^{a_1} \delta
(x_r\!-y_r) \! \ldots \! \delta(x_1\!-y_1) \Theta_\gamma(o,y_1,\! \ldots
\! ,y_r)
\label{multitangent}
\end{equation}
where the $\Theta$ function orders the points of integration along the
loop. The Greek indices represent paired vector and space indices
($\mu_i\equiv a_i x_i$). The holonomy or phase factor associated with a
Lie-algebra valued connection can be written in terms of the
multitangents in the form

\begin{equation}
H_A(\gamma)=\sum_{r=0}^{\infty}\,A_{\mu_1}\ldots A_{\mu_r}\,
X^{\mu_1 \ldots \mu_r}(\gamma)\equiv A_{\b\mu}\,X^{\b\mu}(\gamma)
\label{hol}
\end{equation}
where $\b \mu:=\mu_1\ldots\mu_r$ represents a set of indices of rank
$r(\b \mu)=r$ and $A_{\b \mu}\equiv A_{a_1}(x_1)\cdots A_{a_r}(x_r)$. In
(\ref{hol}) a generalized Einstein convention is assumed (repeated bold
Greek indices indicate a sum of the sets of indices from rank zero to
infinity\footnote{We are using the convention
$A_{\mu_{k+1}\ldots\mu_k}\equiv {\bf 1}$ (the identity matrix) and
$X^{\mu_{k+1}\ldots\mu_k}\equiv1$.}). According to (\ref{hol}), all the
information about loops necessary for the construction of the loop
representation is concentrated into the entire set of multitangent
fields. As far as the loop representation is concerned, the infinite
string of multivector fields can be considered then as totally
equivalent to the loop itself:

\begin{equation}
\gamma \leftrightarrow {\bf X}(\gamma):=\{X^{\b\mu}(\gamma)\,;\;\,
r(\b\mu)=0,\cdots,\infty\}
\end{equation}
Extended loops are generalizations of the multitangent fields to include
more general fields. They have the general form

\begin{equation}
{\bf X}:=\DG_{\b\nu}\,X^{\b\nu}
\end{equation}
where $[{\bf X}]^{\b\mu}\equiv\DG_{\b\nu}^{\b\mu} \,X^{\b\nu}=
X^{\b\mu}$ are now $r(\b\mu)$-times ``contravariant" generalized
multitensors. $\DG_{\b\nu}^{\b\mu}$ acts as a diagonal (infinite
dimensional) identity matrix\footnote{The diagonal elements are given by
the product of generalized delta functions: $\delta^{\mu_1}_{\nu_1}
\cdots \delta^{\mu_r}_{\nu_r}$, being $\delta^{\mu_i}_{\nu_i}:=
\delta^{a_i}_{b_i}\delta(x_i-y_i)$, $\mu_i=a_i x_i$ and $\nu_i=b_i
y_i$.} and the ``covariant" multivectors $\DG_{\b\nu}$ can be viewed as
a canonical basis for the extended loop vectors.

Any (well behaved) application ${\it f}:\{{\bf X}\}\rightarrow {\cal C}
({\cal R})$ from the extended loop space  to complex (or real) numbers
would define a suitable functional of extended loops. In the extended
loop representation the wavefunctions are linear functionals of the
multivector density fields $X^{\mu_1\ldots\mu_r}$:

\begin{equation}
\Psi({\bf X})= \Psi_{\mu_1\ldots\mu_r}\,X^{\mu_1\ldots\mu_r}
\label{lwf}
\end{equation}
The propagators $\Psi_{\mu_1\ldots\mu_r}$ would characterize completely
the linear wavefunction and they could satisfy specific symmetry
requirements. In the case of quantum gravity, the extended loop
wavefunctions can be viewed as formally given by the transform of
wavefunctions with support in the space of connections:

\begin{equation}
\psi({\bf X})=\int DA\,W_{X}[A]\,\psi[A]
\label{ewf}
\end{equation}
where $W_{X}[A]:=Tr[H_A({\bf X})]=Tr[A_{\b\mu}]\,X^{\b\mu}$ is the
extended Wilson functional (an overcomplete basis for the extended loop
transform)\footnote{Notice the close resemblance between the above
formal definitions with those of the conventional loop representation.}.
This means that the propagators of the wavefunctions associated with the
quantum states of gravity can be formally written in the following way

\begin{equation}
\psi_{\mu_1\ldots\mu_r}:=
\int DA\,\psi[A]\,Tr[A_{\mu_1\ldots\mu_r}]
\label{wwf}
\end{equation}
This expression shows that the quantities $\psi_{\mu_1\ldots\mu_r}$
would inherit a set of identities related to the traces of Lie-algebra
matrices (Mandelstam's type identities). For the case of the $SU(2)$
Ashtekar connection we have the following glue-property for the product
of traces:

\begin{equation}
Tr[A_{\b \mu}]\,Tr[A_{\b \nu}]=Tr[A_{\b \mu \b \nu}]+
Tr[A_{\b \mu \bb \nu}]
\label{trA}
\end{equation}
with $\bb \mu\,:=\,(-1)^n\,\b \mu^{-1}$ and $\b
\mu^{-1}\,:=\,\mu_r\ldots\mu_1$. Using the cyclicity of the trace and
(\ref{trA}) it is straightforward to derive the following identities:

\begin{eqnarray}
\psi_{\b \mu}&=&
\psi_{(\b \mu)_c} \label{me1}\\
\psi_{\b \mu}&=&
\psi_{\bb \mu}\label{me2}\\
\psi_{\b \alpha \b \beta \b \pi}+
\psi_{\b \alpha \b \beta \bb \pi}
&=&\psi_{\b \beta \b \alpha \b \pi}
+\psi_{\b \beta \b \alpha \bb \pi}
\label{me3}
\end{eqnarray}
We are going to see that the above set of symmetry properties reduce to
the usual Mandelstam identities for loop wavefunctions when extended
loops are particularized to ordinary loops. The specification to
ordinary loops is made by substituting the general multivector fields by
the multitangent fields:

\begin{equation}
\psi({\bf X}) \rightarrow \psi(\gamma)=
\psi_{\b \mu}\,X^{\b \mu}(\gamma)
\end{equation}
This prescription puts into correspondence any linear extended loop
wavefunction with a loop wavefunction (the converse is not true in
general). The transcription to the case of the conventional loop
representation is then accomplished by using the following group
properties:

\begin{eqnarray}
[X(\gamma_1)\times X(\gamma_2)]^{\b \mu} &:=&
\DG^{\b \mu}_{\b \pi \b \theta} \, X(\gamma_1)^{\b \pi}\,
X(\gamma_2)^{\b \theta}=X^{\b \mu}(\gamma_1\gamma_2)\\
X^{\bb \mu}(\gamma)&=&X^{\b \mu}(\overline{\gamma})
\end{eqnarray}
where $\times$ is the extended group product, $\gamma_1\gamma_2$ is the
group composition in the nonparametric loop space and
$\overline{\gamma}$ is the rerouted loop. Then we can write

\begin{eqnarray}
\psi(\gamma_1\gamma_2)&=&
\psi_{\b\mu}\,
\DG^{(\b \mu)_c}_{\b \alpha \b \beta}\, X^{\b\alpha}(\gamma_1)
X^{\b \beta}(\gamma_2)=\psi(\gamma_2\gamma_1)
\label{m1}\\
\psi(\gamma)
&=&\psi_{\bb \mu}\,X^{\b \mu}(\gamma)=
\psi_{\b \mu}\,X^{\bb \mu}(\gamma)=
\psi(\overline{\gamma})\label{m2}\\
\psi(\gamma_1\gamma_2\gamma_3)&=&
\psi_{\b \alpha \b \beta \b \pi}\,
X^{\b \alpha}(\gamma_1)X^{\b \beta}(\gamma_2)X^{\b \pi}(\gamma_3)
\nonumber\\
&=&[\psi_{\b \beta \b \alpha \b \pi}+
\psi_{\b \beta \b \alpha \bb \pi}
-\psi_{\b \alpha \b \beta \bb \pi}]
X^{\b \alpha}(\gamma_1)X^{\b \beta}(\gamma_2)X^{\b \pi}(\gamma_3)
\nonumber\\
&=&\psi(\gamma_2\gamma_1\gamma_3)+
\psi(\gamma_2\gamma_1\overline{\gamma_3})
-\psi(\gamma_1\gamma_2\overline{\gamma_3})
\label{m3}
\end{eqnarray}
Equations (\ref{m1})-(\ref{m3}) are the usual Mandelstam identities for
loop wavefunctions. As we have shown they can be viewed as particular
cases of the symmetry identities (\ref{me1})-(\ref{me3}).

We review now the transformation properties of the extended loop
wavefunctions under diffeomorphisms\footnote{The properties of extended
objects under general coordinate transformations were first considered
by \cite{GaLe} for multitangent fields and later by \cite{DiGaGr1} for
extended loops.}. Extended loops behave as multivector densities and the
linear extended loop wavefunctions change under infinitesimal coordinate
transformations $x`^a=x^a + \eta^a (x)$ in the way

\begin{equation}
\psi({\bf X}')=\psi({\bf X})+\eta^{ax}\,{\cal C}_{ax}\,\psi({\bf X})
\label{difwf}
\end{equation}
where

\begin{equation}
{\cal C}_{ax}:=[{\cal F}_{ab}(x)\times{\bf X}^{(bx)}]^{\b
\mu}\frac{\delta}{\delta\,X^{\b \mu}}\equiv{\cal F}_{ab}^{\b
\alpha}(x)\,X^{(bx\,\b \beta)_c}\frac{\delta}{\delta\,X^{\b\alpha\b\beta}}
\label{exdif}
\end{equation}
This result can be derived from the generator of diffeomorphisms in the
connection space (via the extended loop transform) \cite{DiGaGr2} or by
using the explicit transformation properties of extended loops under
infinitesimal coordinate transformations \cite{Di}. In (\ref{exdif}),
$[{\bf X}^{(bx)}]^{\b\beta}:=X^{(bx\,\b\beta)_c}$ is the one-point-X
($c$ indicates the average under cyclic permutation) and

\begin{equation}
{\cal F}_{ab}^{\alpha_1\ldots\alpha_r}(x):=\epsilon_{abc}\,[-
\DG^{\alpha_1\ldots\alpha_r}_{\nu_1}\,g^{cx\,\nu_1}+
\DG^{\alpha_1\ldots\alpha_r}_{\nu_1\nu_2}\,\epsilon^{cx\,\nu_1\nu_2}]
\label{fab}
\end{equation}
$g^{cx\,\nu_1}$ is the inverse of the two point propagator of the
Chern-Simons theory,

\begin{equation}
g^{cx\,\nu_1}:=\epsilon^{cb_1k}\,\partial_k\,\delta(x-y_1)
\end{equation}
and

\begin{equation}
\epsilon^{cx\,\nu_1\nu_2}:=\epsilon^{cb_1b_2}
\delta(x-y_1)\delta(x-y_2)
\end{equation}
In the last expressions a mixed notation of Greek and paired indices
were used. Let us consider now the specification of (\ref{difwf}) to the
case of ordinary loops. From (\ref{lwf}) and (\ref{exdif}) we can write

\begin{eqnarray}
{\cal C}_{ax}\,\psi[{\bf X}(\gamma)]&=&\psi_{\b\alpha\b\pi\b\theta}
\,{\cal F}_{ab}^{\b\alpha}(x)\,X^{\b\theta\,bx\,\b\pi}(\gamma)
\nonumber\\
&=&\oint_{\gamma}\,dy^b \delta(x-y)\psi_{\b\alpha\b\pi\b\theta}
\,{\cal F}_{ab}^{\b\alpha}(y)\,
X^{\b\pi}(\gamma_y^o)\,X^{\b\theta}(\gamma_o^y)
\nonumber\\
&=&\oint_{\gamma}\,dy^b \delta(x-y)\psi_{\b\mu}
\,[{\cal F}_{ab}(x)\times{\bf X}(\gamma_y^o)\times{\bf
X}(\gamma_o^y)]^{\b\mu}
\end{eqnarray}
where $\gamma_o^y$ is an open path form $o$ to $y$. Introducing now the
group identity $1={\bf X}(\overline{\gamma}_o^y)\times {\bf
X}(\gamma_o^y)$ and using the cyclicity of the set of indices $\b \mu$
we get

\begin{eqnarray}
{\cal C}_{ax}\psi(\gamma)&=&
\oint_{\gamma}\,dy^b \delta(x-y)\psi_{\b\mu}
[{\bf X}(\gamma_o^y)\times{\cal F}_{ab}(y)\times
{\bf X}(\overline{\gamma}_o^y)\times{\bf X}(\gamma_o^y)\times
{\bf X}(\gamma_y^o)]^{\b\mu}
\nonumber\\
&\equiv&
\oint_{\gamma}\,dy^b\, \delta(x-y)\,\Delta_{ab}(\gamma^y)
\,\psi(\gamma)
\label{diflw}
\end{eqnarray}
where

\begin{equation}
\Delta_{ab}(\gamma^y)\,{\bf X}(\gamma):=
[{\bf X}(\gamma_o^y)\times{\cal F}_{ab}(y)\times
{\bf X}(\overline{\gamma}_o^y)]\times{\bf X}(\gamma)
\end{equation}
is the loop derivative in the non-parametric loop space \cite{GaTr}. The
expression (\ref{diflw}) gives the correct transformation law of loop
wavefunctions under diffeomorphisms.

\section{Extended knot invariants}
The previous discussion shows that the solutions of the equation

\begin{equation}
{\cal C}_{ax}\,\psi({\bf X}) = 0
\label{eki}
\end{equation}
can be viewed as defining ``extended knot invariants" (in the sense that
the restriction of the domain of definition of the solution to ordinary
loops will always gives a knot invariant). The analysis of this equation
in the extended loop space reveals the existence of a specific mechanism
linking the action of the operator with some definite structure of the
propagators $\psi_{\b\mu}$. To see this we write (\ref{eki}) in the form

\begin{equation}
{\cal C}_{ax}\,\psi({\bf X}) =
\sum_{r=0}^{\infty} \epsilon_{abc}\psi_{\mu_1 \ldots \mu_r}[-
g^{cx\,\mu_1}\,X^{(bx\,\mu_2 \ldots \mu_r)_c}+
\epsilon^{cx\,\mu_1 \mu_2}\,X^{(bx\,\mu_3 \ldots \mu_r)_c}]=0
\label{eki2}
\end{equation}
The possibility to systematize the search of solutions of (\ref{eki2}) is
based on the following two punctuations:

\begin{description}
\item{P1:} The propagators $\psi_{\mu_1\ldots\mu_r}$ are expressed
completely in terms of the two ($g_{\cdot\cdot}$) and three
($h_{\cdot\cdot\cdot}$) point propagators of the Chern-Simons theory.

\item{P2:} The propagators $\psi_{\mu_1\ldots\mu_r}$ exhibit a definite
skeleton structure in terms of the basic ``blocks" $g$ and $h$.

\end{description}
The architectonic profile of the skeleton is:

\begin{description}
\item{S1:} The propagators {\it vanish} for a certain (maximum) rank.
That is to say, $\psi_{\mu_1\ldots\mu_r}=0$ for all $r>N$.

\item{S2:} For $r=N$, $\psi_{\mu_1\ldots\mu_N}$ is expressed entirely in
terms of the two point Chern-Simons propagator. Specifically, they are
cyclic combinations of products of $g$'s.

\item{S3:} For the successive decreasing ranks the ``free" $g$'s are
progressively substituted by ``connected" $h$'s. The final expression is
always cyclic in the covariant Greek indices.

\item{S4:} The minimum rank $n$ does not contain any free two point
Chern-Simons propagator.

\end{description}
The above nomenclature of free-$g$ and connected-$h$ refers to the
analytic properties of the Chern-Simons propagators. In fact, the three
point propagator $h_{\mu_1\mu_2\mu_3}$ is nothing else than the
``contraction" of three free two point $g_{\mu_i\alpha_i}$ with the
``vertex" $\epsilon^{\alpha_1\alpha_2\alpha_3}$:

\begin{equation}
h_{\mu_1\mu_2\mu_3}:=\epsilon^{\alpha_1\alpha_2\alpha_3}\,
g_{\mu_1\alpha_1}g_{\mu_2\alpha_2}g_{\mu_3\alpha_3}
\label{h}
\end{equation}
The free propagator can be written in the following way,

\begin{equation}
g_{\mu_1\mu_2}\equiv\epsilon_{a_1 a_2 c}\phi^{\,cx_1}_{x_2}:=
-\epsilon_{a_1 a_2 c}\frac{\partial^{\,c}}{\nabla^2}\delta(x_1-x_2)
\label{g}
\end{equation}
The function $\phi$ plays a prominent role in the extended loop
formalism. It fixes a transverse prescription for the multivector
fields through the following decomposition of the identity:
$\delta^{\mu_1}_{\nu_1}=
\delta{^{\phantom{T}\mu_1}_{\mbox{\scriptsize{T}}\phantom{ax}
\nu_1}}+\phi^{\,\mu_1}_{\,z_1,\,d_1}$. Then

\begin{equation}
Y^{\mu_1}= \delta{^{\phantom{T}\mu_1}_{\mbox{\scriptsize{T}}\phantom{ax}
\nu_1}}\,X^{\nu_1}
\label{Ytransverso}
\end{equation}
is a transverse (divergence free) field. This property can be directly
generalized to the case of multivector fields \cite{DiGaGr1}. Notice that

\begin{equation}
{\cal F}_{ab}^{\mu_1}\,g_{\mu_1\mu_2}=-
\epsilon_{abc}\,\delta{^{\phantom{T}cx}_{\mbox{\scriptsize{T}}\phantom{ax}
\mu_2}}
\label{F1g}
\end{equation}
whereas

\begin{equation}
{\cal F}_{ab}^{\mu_1}\,h_{\mu_1\mu_2\mu_3}=-
\epsilon_{abc}\,\delta{^{\phantom{T}cx}_{\mbox{\scriptsize{T}}\phantom{ax}
\alpha_1}}\epsilon^{\alpha_1\alpha_2\alpha_3}\,
g_{\mu_2\alpha_2}g_{\mu_3\alpha_3}
\label{F1h}
\end{equation}
with

\begin{equation}
\delta{^{\phantom{T}cx}_{\mbox{\scriptsize{T}}\phantom{ax}
\alpha_1}}\epsilon^{\alpha_1\alpha_2\alpha_3}=
\epsilon^{cx\,\alpha_2\alpha_3}+(\phi^{\,cx}_{z_2}-\phi^{\,cx}_{z_3})
g^{\alpha_2\alpha_3}
\label{deltaepsilon}
\end{equation}
being $z_k$ the spatial part of the indices $\alpha_k$. Let us return to
the skeleton rules S1-S4. They mirror the analytic properties of knot
invariants known at present\footnote{The state given by the exponential
of the self-linking number is an exception to S1. The breakdown of S1
(i.e. the appearance of series) introduces convergence problems into the
formalism. The consideration of this question is out of the scope of the
present discussion.}. A point to stress is that no other propagators
than the Chern-Simons are known to participate in the analytic
properties of knots. In our case the rules S1-S4 provide the first
ingredient for a systematization: the propagators
$\psi_{\mu_1\ldots\mu_r}$ are structured according a sequential
arrangement of some basic blocks in the form

\begin{equation}
\psi({\bf X})=g_{\cdot\cdot}g_{\cdot\cdot}\cdots g_{\cdot\cdot}\,
X^{\cdots\cdots\cdots}+h_{\cdot\cdot\cdot}\cdots g_{\cdot\cdot}\,
X^{\cdot\cdot\cdots\cdots}+\cdots+
h_{\cdot\cdot\cdot}\cdots h_{\cdot\cdot\cdot}\,
X^{\cdots\cdots}
\label{gwf}
\end{equation}
The second ingredient appears when (\ref{eki2}) (and in particular
(\ref{F1g}) and (\ref{F1h})) start to work. The following general
observations can be outlined from the action of the diffeomorphism
operator over wavefunctions of the type (\ref{gwf}):

\begin{description}
\item{O1:} The successive ranks of the wavefunction are linked
intimately. As a general rule, ${\cal C}_{ax}$ acting on a generic rank
$r$ produces two types of contributions, one of rank $r$ and other of
rank $r-1$. For $r>n$, the contribution of rank $r-1$ is always canceled
by terms that appear when the operator act on the rank $r-1$ of the
wavefunction. A chain of cancellations is then induced by the action of
the diffeomorphism operator\footnote{Exactly the same chain of
cancellations takes place in the case of the Hamiltonian. This is a
characteristic of the extended loop representation, see \cite{Gr6}.}.

\item{O2:} For $n<r\leq N$ there exist in general remnant terms that do
not enter into the chain of cancellations. The general form of these
terms is $\epsilon_{abc}\psi_{\ldots\ldots} X^{(bx\,\ldots\,cx
\,\ldots)_c}$ and they have to vanish by means of symmetry
considerations\footnote{For extended knots this contribution involves in
general a symmetric expression in the indices $bx$ and $cx$.}.

\item{O3:} The term of lower rank $n$ is responsible of the closure of
the chain in a consistent way.

\end{description}
In what follows we are going to use the above punctuations as guidelines
to establish a systematic method to construct extended knot invariants.

\section{Extended knot families}
We start by describing in general the properties of the final product we
get: the extended knot families. They are given by sets $\{\psi_i\}^N_n$
of linear extended loop wavefunctions with the same maximum and minimum
ranks that satisfy the following conditions:

\begin{description}
\item{F1:} ${\cal C}_{ax}\,\psi_i=0$ for all $i$.

\item{F2:} $\psi_i({\bf X})=\psi^{i}{_{\b\mu}}\,X^{\b\mu}$, with
$\psi^{i}{_{\b\mu}}$ a cyclic propagator that fulfill the skeleton rules
S1-S4.

\item{F3:} $\psi^{i}{_{\mu_1\ldots\mu_n}}$ is the {\it same} for all
members of the family.

\end{description}

\subsection{The maximum (even) rank}
For $r=N$ the propagators $\psi^i{_{\mu_1\ldots\mu_N}}$ are written as
cyclic combinations of products of $g$'s. For $N=6$ we have for example
the following possibilities:

\begin{eqnarray}
C^1{_{\mu_1\ldots\mu_6}}&=&g_{\mu_1\mu_4}g_{\mu_2\mu_5}g_{\mu_3\mu_6}\\
C^2{_{\mu_1\ldots\mu_6}}&=&g_{\mu_1\mu_2}g_{\mu_3\mu_4}g_{\mu_5\mu_6}+
g_{\mu_1\mu_6}g_{\mu_2\mu_3}g_{\mu_4\mu_5}\\
C^3{_{\mu_1\ldots\mu_6}}&=&g_{\mu_1\mu_2}g_{\mu_3\mu_6}g_{\mu_4\mu_5}+
g_{\mu_1\mu_4}g_{\mu_2\mu_3}g_{\mu_5\mu_6}+
g_{\mu_1\mu_6}g_{\mu_2\mu_5}g_{\mu_3\mu_4}\\
C^4{_{\mu_1\ldots\mu_6}}&=&g_{\mu_1\mu_3}g_{\mu_2\mu_5}g_{\mu_4\mu_6}+
g_{\mu_1\mu_4}g_{\mu_2\mu_6}g_{\mu_3\mu_5}+
g_{\mu_1\mu_5}g_{\mu_2\mu_4}g_{\mu_3\mu_6}\\
C^5{_{\mu_1\ldots\mu_6}}&=&g_{\mu_1\mu_2}g_{\mu_3\mu_5}g_{\mu_4\mu_6}+
g_{\mu_1\mu_3}(g_{\mu_2\mu_6}g_{\mu_4\mu_5}+g_{\mu_2\mu_4}g_{\mu_5\mu_6})
+\nonumber\\
&&g_{\mu_1\mu_6}g_{\mu_2\mu_4}g_{\mu_3\mu_5}+
g_{\mu_1\mu_5}(g_{\mu_2\mu_3}g_{\mu_4\mu_6}+g_{\mu_2\mu_6}g_{\mu_3\mu_4})
\end{eqnarray}
We need some suitable notation in order to reflect appropriately the
action of the operator ${\cal C}_{ax}$. Putting

\begin{equation}
\psi^i{_{\mu_1\ldots\mu_N}}=\sum_{[k]}g_{\mu_1\mu_k}
G^{i}{_{\mu_2\ldots\mu_{k-1}\mu_{k+1}\ldots\mu_N}}
\label{max1}
\end{equation}
we found

\begin{eqnarray}
\lefteqn{ {\cal C}_{ax}\psi_{i}\mid_{r=N}=-\sum_{[k]}
\epsilon_{abc}\delta{^{\phantom{T}cx}_{\mbox{\scriptsize{T}}\phantom{ax}
\mu_k}}G^{i}{_{\mu_2\ldots\mu_{k-1}
\mu_{k+1}\ldots\mu_N}}
X^{(bx\,\mu_2\ldots\mu_N)_c} }
\nonumber\\
&&+\sum_{[k,l]}g_{\mu_k\,[ax}g_{bx]\,\mu_l}
G^{i}{_{\mu_3\ldots\mu_{k-1}\mu_{k+1}\ldots\mu_{l-1}\mu_{l+1}\ldots\mu_N}}
X^{(bx\,\mu_3\ldots\mu_N)_c}
\label{max2}
\end{eqnarray}
Using the fact that $\delta_{\mbox{\scriptsize{T}}}=\delta-\phi$ and
integrating by parts the longitudinal projector contribution we obtain

\begin{eqnarray}
\lefteqn{ {\cal C}_{ax}\psi_{i}\mid_{r=N}=-\sum_{[k]}
\epsilon_{abc}G^{i}{_{\mu_2\ldots\mu_{k-1}\mu_{k+1}\ldots\mu_N}}
X^{(bx\,\mu_2\ldots\mu_{k-1}\,cx\,\mu_{k+1}\ldots\mu_N)_c} }
\nonumber\\
&&-\sum_{[k]}\epsilon_{abc}
(\phi^{\,cx}_{\,x_{k-1}}-\phi^{\,cx}_{\,x_{k+1}})
G^{i}{_{\mu_2\ldots\mu_{k-1}\mu_{k+1}\ldots\mu_N}}\,
X^{(bx\,\mu_2\ldots\mu_{k-1}\mu_{k+1}\ldots\mu_N)_c}
\nonumber\\
&&+\sum_{[k,l]}g_{\mu_k [ax}\,g_{\,bx] \,\mu_l}
G^{i}{_{\mu_3\ldots\mu_{k-1}\mu_{k+1}\ldots\mu_{l-1}\mu_{l+1}\ldots\mu_N}}
X^{(bx\,\mu_3\ldots\mu_N)_c}
\label{rN}
\end{eqnarray}
where we have used the differential constraint \cite{DiGaGr1} of the
multivector fields. We get three contributions, the first of rank $N$
and the others of rank $N-1$. It is easy to see that a sufficient
condition for the contribution of rank $N$ to vanish is the cyclicity of
$\psi_{\mu_1\ldots\mu_N}$\footnote{I thank C. Di Bartolo to point me out
this fact.}.

\subsection{The rank $N-1$}
The next rank is formed by substituting two $g$'s for one
$h\,$\footnote{There is not a unique way to make this passage guided
only by S3. For example for $N=10$ we have two possibilities:
$h_{\mu_1\mu_2\mu_3}g_{\mu_4\mu_5}g_{\mu_6\mu_7}g_{\mu_8\mu_9}$ or
$h_{\mu_1\mu_2\mu_3}h_{\mu_4\mu_5\mu_6}h_{\mu_7\mu_8\mu_9}$. In order to
simplify the discussion we shall limit to consider only the first
case.}. For $N-1=5$ one finds two cyclic combinations of this type

\begin{eqnarray}
C^1{_{\mu_1\ldots\mu_5}}&=&g_{(\mu_1\mu_2}h_{\mu_3\mu_4\mu_5)_c}
\\
C^2{_{\mu_1\ldots\mu_5}}&=&g_{(\mu_1\mu_3}h_{\mu_2\mu_4\mu_5)_c}
\end{eqnarray}
Now we have to distinguish when $\mu_1$ belongs to a $g$ or to an $h$.
Writing

\begin{eqnarray}
\lefteqn{ \psi^i{_{\mu_1\ldots\mu_{N-1}}}=
\sum_{[k]}g_{\mu_1\mu_k}
H^{i}{_{\mu_2\ldots\mu_{k-1}\mu_{k+1}\ldots\mu_{N-1}}} }
\nonumber\\
&&+\sum_{[k,l]}h_{\mu_1\mu_k\mu_l}
G^{i}{_{\mu_2\ldots\mu_{k-1}\mu_{k+1}\ldots\mu_{l-1}\mu_{l+1}
\ldots\mu_{N-1}}}
\end{eqnarray}
we get

\begin{eqnarray}
\hspace{-0.7cm} \lefteqn{
{\cal C}_{ax}\psi_{i}\mid_{r=N-1}=-\sum_{[k]}
\epsilon_{abc}H^{i}{_{\mu_2\ldots\mu_{k-1}
\mu_{k+1}\ldots\mu_{N-1}}}
X^{(bx\,\mu_2\ldots\mu_{k-1}\,cx\,\mu_{k+1}\ldots\mu_{N-1})_c} }
\nonumber\\
&&-\sum_{[k,l]}
\epsilon_{abc}
(\phi^{\,cx}_{\,x_l}-\phi^{\,cx}_{\,x_k})g_{\mu_k\mu_l}\,
G^{i}{_{\mu_3\ldots\mu_{k-1}\mu_{k+1}\ldots\mu_{l-1}\mu_{l+1}\ldots\mu_N}}
X^{(bx\,\mu_3\ldots\mu_N)_c}
\nonumber\\
&&-\sum_{[k,l]}
g_{\mu_k [ax}\,g_{\,bx] \,\mu_l}\,
G^{i}{_{\mu_3\ldots\mu_{k-1}\mu_{k+1}\ldots\mu_{l-1}\mu_{l+1}\ldots\mu_N}}
X^{(bx\,\mu_3\ldots\mu_N)_c}
\nonumber\\
&&+\,\mbox{\footnotesize{terms of rank $N-2$}}
\label{rN-1}
\end{eqnarray}
The comparison of (\ref{rN}) and (\ref{rN-1}) permits to conclude that:
1) the terms of the type described in O2 appear whenever the $\mu_1$
index lies in a free-$g$. 2) The presence of the $h$ in the rank $N-1$
of the wavefunction is responsible for the generation of terms with
exactly the form needed to cancel {\it all} the contribution of rank
$N-1$ in (\ref{rN}). In other worlds, the connected-$h$ starts the chain
of cancellations associated with the diffeomorphism operator.

As in the case of rank $N$, the cyclicity of
$\psi_{\mu_1\ldots\mu_{N-1}}$ assures the annulment of the first sum in
(\ref{rN-1}). Then, {\it if} the wavefunction's propagators of ranks $N$
and $N-1$ are suitables to produce the cancellation of the remaining sums
in (\ref{rN}) and (\ref{rN-1}) between themselves we can write

\begin{equation}
{\cal C}_{ax}\,\{\psi_{i}\mid_{r=N}+\psi_{i}\mid_{r=N-1}\}=
\mbox{\footnotesize{terms of rank $N-2$}}
\end{equation}
Due to the higher content of connected propagators as well as the
multiplication of possibilities to go from one rank to the following,
the analysis becomes more intricate as one advances along the chain. The
main noticed characteristics are however preserved, so one can write in
general

\begin{equation}
{\cal C}_{ax}\,\{\psi_{i}\mid_{r=N}+\psi_{i}\mid_{r=N-1}+\cdots
+\psi_{i}\mid_{r=m}\}=
\mbox{\footnotesize{terms of rank $m-1$}}
\label{chain}
\end{equation}

\subsection{The minimum rank}
What happens in (\ref{chain}) when one gets the minimum rank? The only
possibility is that the contribution of rank $n-1$ vanishes identically.
This condition has a definite meaning that we clarify with a simple
example. The minimum rank is composed only by connected propagators.
The most simple case is a single $h$. One finds for this case

\begin{eqnarray}
{\cal C}_{ax}h_{\mu_1\mu_2\mu_3}X^{\mu_1\mu_2\mu_3}&=&
\{-g_{\mu_1\,[ax\,}g_{bx]\,\mu_2}+\epsilon_{abc}(\phi^{\,cx}_{\,x_1}
-\phi^{\,cx}_{\,x_2})g_{\mu_1\mu_2}\}\,X^{(bx\,\mu_1\mu_2)_c}
\nonumber\\
&&+2\{h_{ax\,bx\,\mu_1}-\epsilon_{abc}\phi^{\,cx}_{\,z}\phi^{\,dz}_{\,x}
g_{\mu_1\,dz}\}\,X^{(bx\,\mu_1)_c}
\label{difh}
\end{eqnarray}
where an integration in the spatial variable $z$ is assumed. Notice that
we can not use here symmetry arguments of the type O2. The only general
way to annul the contribution of rank $2$ in (\ref{difh}) is by
demanding

\begin{equation}
\{h_{ax\,bx\,\mu_1}-\epsilon_{abc}\phi^{\,cx}_{\,z}\phi^{\,dz}_{\,x}
g_{\mu_1\,dz}\}\equiv 0
\end{equation}
But this is a formal identity for the propagators! Indeed, from
(\ref{h}) it is straightforward to see that the second term of the
l.h.s. is a formal equivalent way to write the three point propagator
with two spatial indices fixed at $x$. The annulment of the contribution
of rank $n-1$ for $m=n$ in (\ref{chain}) appears then as a consistency
condition for the closure of the chain.

\subsection{The method}
We have seen that the existence of a systematic procedure to obtain
diffeomorphism invariants is suggested by the characteristics of the
extended loop calculus. Also that the complexity of the analysis grows
significantly at higher ranks. A practical way to proceed is the
following:

\vspace{0.3cm}
\noindent
{\it a)} Select a cyclic expression of Chern-Simons
propagators that not contain any free $g$. Verify the consistence
condition $O3$ according the analysis of the preceding subsection. We
get then a candidate for the minimum rank $n$.

\vspace{0.2cm}
\noindent
{\it b)} Find the expressions of rank $n\!\!+\!\!1$ that make ${\cal
C}_{ax}\{\psi_{i}\!\mid_{r={n+1}}+\psi_{i}\!\mid_{r=n}\}=
\mbox{\footnotesize{terms of rank $n+1$}}$. Verify that the terms of the
form $\epsilon_{abc}\psi_{\ldots\ldots}X^{(bx\,\ldots\,cx\,\ldots)_c}$
vanish by symmetry considerations.

\vspace{0.3cm}
\noindent
{\it c)} Repeat the procedure until all the connected parts are
substituted by free-$g$'s.

\vspace{0.3cm}
\noindent
By this procedure the extended knot families are generated.

\section{The family $\{\psi_i\}^6_4$}
We apply now the method to construct a family with maximum rank $6$ and
minimum rank $4$. For $n=4$

\begin{equation}
C_{\mu_1\ldots\mu_4}=
h_{\mu_1\mu_2\alpha}g^{\alpha\beta}h_{\beta\mu_3\mu_4}-
h_{\mu_1\mu_4\alpha}g^{\alpha\beta}h_{\beta\mu_2\mu_3}
\end{equation}
is a suitable expression that not contains free-$g$'s\footnote{Another
more connected possibility is:
$h_{\mu_1\alpha_1\alpha_2}g^{\alpha_1\beta_1}g^{\alpha_2\beta_2}
h_{\beta_1\beta_2\mu_2}h_{\mu_3\alpha_3\alpha_4}g^{\alpha_3\beta_3}
g^{\alpha_4\beta_4}h_{\beta_3\beta_4\mu_4}+ \mbox{cyclic
permutations}$.}. We get for this case (see \cite{Gr6}):

\begin{eqnarray}
\lefteqn{
\hspace{-0.3cm}
{\cal C}_{ax}\,C_{\mu_1\ldots\mu_4}\,X^{\mu_1\ldots\mu_4}=
2\{g_{\mu_1[\,ax}h_{bx\,]\,\mu_2 \mu_3}
-\epsilon_{abc}\phi^{\,cx}_{\,x_1}h_{\mu_1\mu_2\mu_3}
}\nonumber\\
&&\hspace{-0.3cm}+\epsilon_{abc}\phi^{\,cx}_{\,z}
(\phi^{\,dz}_{\,x_3}-\phi^{\,dz}_{\,x_2})g_{\mu_1 \,dz}g_{\mu_2\mu_3}
\} X^{(bx\,\mu_1 \mu_2 \mu_3)_c}
+{\cal I}_{ax\,bx\,\mu_1\mu_2}
\,X^{(bx\,\mu_1\mu_2)_c}
\label{Cr4}
\end{eqnarray}
with
\begin{eqnarray}
&&\hspace{-0.6cm}{\cal I}_{ax\,bx\,\mu_1\mu_2}:=
-2h_{ax\,bx\,\alpha} g^{\alpha\beta}h_{\mu_1\mu_2\beta} +
h_{\mu_2\alpha\,[\,ax} h_{bx\,]\,\mu_1\beta}g^{\alpha\beta}
\nonumber\\
&&\hspace{-0.3cm}
-2\epsilon_{abc}\phi^{\,cx}_{\,z}(\phi^{\,dz}_{\,x}-
\phi^{\,dz}_{\,x_1})h_{dz\,\mu_1\mu_2}
-2\epsilon_{abc}\phi^{\,cx}_{\,z}\phi^{\,dz}_{\,y}
(\phi^{\,ey}_{\,x}+\phi^{\,ey}_{\,x_1})g_{\mu_1\,dz}g_{\mu_2\,ey}
\end{eqnarray}
In reference \cite{Gr6} it was demonstrated that formally ${\cal
I}_{ax\,bx\,\mu_1\mu_2}\equiv0$, so the consistency
condition is fulfilled. For the rank $n+1=5$ the propagators have the
general form $g_{(\cdot\cdot}h_{\cdot\cdot\cdot)_c}$. According the
results listed in the Appendix it is easy to see that the combinations
$C^{1}_{\mu_1\ldots\mu_5}X^{\mu_1\ldots\mu_5}-
C_{\mu_1\ldots\mu_4}X^{\mu_1\ldots\mu_4}$ and
$C^{2}_{\mu_1\ldots\mu_5}X^{\mu_1\ldots\mu_5}+
C_{\mu_1\ldots\mu_4}X^{\mu_1\ldots\mu_4}$ satisfy the requirement {\it
b)}. The next rank would involve expressions of the form
$g_{(\cdot\cdot}g_{\cdot\cdot}g_{\cdot\cdot)_c}$ and the procedure is
completed. By using again the results of the Appendix one finds:

\begin{eqnarray}
\psi_1&=&[C^{2}_{\mu_1\ldots\mu_6}+C^{3}_{\mu_1\ldots\mu_6}]
X^{\mu_1\ldots\mu_6}-C^{1}_{\mu_1\ldots\mu_5}X^{\mu_1\ldots\mu_5}+
C_{\mu_1\ldots\mu_4}X^{\mu_1\ldots\mu_4}\label{1}\\
\psi_2&=&[C^{1}_{\mu_1\ldots\mu_6}+C^{4}_{\mu_1\ldots\mu_6}+
C^{5}_{\mu_1\ldots\mu_6}]X^{\mu_1\ldots\mu_6}+
C^{1}_{\mu_1\ldots\mu_5}X^{\mu_1\ldots\mu_5}
\nonumber\\
&&-C_{\mu_1\ldots\mu_4}X^{\mu_1\ldots\mu_4}\label{2}\\
\psi_3&=&[2C^{1}_{\mu_1\ldots\mu_6}+C^{4}_{\mu_1\ldots\mu_6}]
X^{\mu_1\ldots\mu_6}+C^{2}_{\mu_1\ldots\mu_5}X^{\mu_1\ldots\mu_5}+
C_{\mu_1\ldots\mu_4}X^{\mu_1\ldots\mu_4}\label{3}\\
\psi_4&=&[-C^{1}_{\mu_1\ldots\mu_6}+C^{2}_{\mu_1\ldots\mu_6}
+C^{3}_{\mu_1\ldots\mu_6}+C^{5}_{\mu_1\ldots\mu_6}]
X^{\mu_1\ldots\mu_6}-C^{2}_{\mu_1\ldots\mu_5}X^{\mu_1\ldots\mu_5}
\nonumber\\
&&-C_{\mu_1\ldots\mu_4}X^{\mu_1\ldots\mu_4}\label{4}
\end{eqnarray}
No other analytic expressions of the general form
$\psi=g_{\cdot\cdot}g_{\cdot\cdot}g_{\cdot\cdot}
X^{\cdot\cdot\cdot\cdot\cdot\cdot}+g_{\cdot\cdot}h_{\cdot\cdot\cdot}
X^{\cdot\cdot\cdot\cdot\cdot}+h_{\cdot \cdot \star}g^{\star \star}
h_{\star \cdot \cdot}X^{\cdot \cdot \cdot \cdot}$ invariant under
diffeomorphisms apart form that listed above would exist. Two immediate
questions arise: $1)$ to which knot invariants would correspond the
different members of the family $\{\psi_i\}^6_4$, and $2)$ which among
them would satisfy the Mandelstam identities?

To answer the first question we need to introduce another player into
scene: the $*$-product of diffeomorphism invariants \cite{Gr5,Grprep}.
The $*$-product allows to put into correspondence linear invariant
expressions with the usual product of diffeomorphism invariants. For
example, the $*$-square of the Gauss invariant
$\varphi_G:=g_{\mu_1\mu_2}X^{\mu_1\mu_2}$ is defined in the following
way\footnote{The underline of an ordered subset
$\underline{\mu_1\ldots\mu_k}\mu_{k+1}\ldots\mu_r$ of $k$ indices
indicates in general the linear combination of multivectors obtained by
permuting the indices in all possible ways but preserving the relative
order of the subsets $\mu_1\ldots\mu_k$ and $\mu_{k+1}\ldots\mu_r$ among
themselves.}

\begin{eqnarray}
(*\,\varphi_G)^2&:=&g_{\mu_1\mu_2}g_{\mu_3\mu_4}
X^{\underline{\mu_1\mu_2}\mu_3\mu_4}
\nonumber\\
&=&g_{\mu_1\mu_2}g_{\mu_3\mu_4}(X^{\mu_1\mu_2\mu_3\mu_4}+
X^{\mu_1\mu_3\mu_2\mu_4}+
X^{\mu_1\mu_3\mu_4\mu_2}
\nonumber\\
&&+X^{\mu_3\mu_1\mu_2\mu_4}+X^{\mu_3\mu_1\mu_4\mu_2}+
X^{\mu_3\mu_4\mu_1\mu_2})
\nonumber\\
&\equiv&2(g_{\mu_1\mu_2}g_{\mu_3\mu_4}+g_{\mu_1\mu_3}g_{\mu_2\mu_4}
+g_{\mu_1\mu_4}g_{\mu_2\mu_3})X^{\mu_1\mu_2\mu_3\mu_4}
\label{*square}
\end{eqnarray}
For those extended loops that satisfy the {\it algebraic constraint}
$X^{\underline{\mu_1\mu_2}\mu_3\mu_4}= X^{\mu_1\mu_2}X^{\mu_3\mu_4}$,
(\ref{*square}) reduces to the usual product of Gauss invariants (this
is a general property of the $*$-composition law). In general, the
$*$-product of two diffeomorphism invariants is a diffeomorphism
invariant \cite{Di}. From equations (\ref{1})-(\ref{4}) the following
relationships (constraints) between the members of the family are found:

\begin{eqnarray}
\psi_1 +\psi_2=\psi_3 +\psi_4 &=&\textstyle{1\over3!}(*\,\varphi_G)^3
\label{i1}\\
\psi_2 +\psi_3&=& \varphi_G * J_2\label{i2}
\end{eqnarray}
where $J_2:=g_{\mu_1\mu_3}g_{\mu_2\mu_4}X^{\mu_1\mu_2\mu_3\mu_4}
+h_{\mu_1\mu_2\mu_3}X^{\mu_1\mu_2\mu_3}$ is the analytic part of the
second coefficient of the Alexander-Conway knot polynomial (that
coincides with the second coefficient of certain expansion of the Jones
polynomial). We conclude that of the four members of the family, only
one would represent a new diffeomorphism invariant. Taking this single
invariant as $\psi_3\equiv J_3$, we get from (\ref{i1})-(\ref{i2})

\begin{eqnarray}
\psi_1  &=&\textstyle{1\over3!}(*\,\varphi_G)^3-
\varphi_G * J_2 + J_3 \\
\psi_2 &=& \varphi_G * J_2 - J_3 \\
\psi_3&=& J_3 \\
\psi_4 &=&\textstyle{1\over3!}(*\,\varphi_G)^3 - J_3
\end{eqnarray}
$\varphi_G$ and $J_2$ belong to lower rank families \cite{Gr9}:

\begin{eqnarray}
&&\{\psi_i\}^2_2 = \{ \varphi_G \} \\
&&\{\psi_i\}^4_3 = \{J_2\,,\;{\textstyle{1\over2}}(*\,\varphi_G)^2 -J_2\}
\end{eqnarray}
Using the above results the family $\{\psi_i\}^6_4$ can be written in
the form

\begin{equation}
\{\psi_i\}^6_4=\{K_3\,,\;\textstyle{1\over3!}(*\,\varphi_G)^3-
K_3\,,\;J_3\,,\;
\textstyle{1\over3!}(*\,\varphi_G)^3-J_3\}
\label{family}
\end{equation}
with

\begin{equation}
K_3:=J_3- J_2 * \varphi_G +\textstyle{1\over3!}(*\,\varphi_G)^3
\label{k3j3}
\end{equation}
$K_3$ is related to the third coefficient of certain expansion of the
Kauffman bracket knot polynomial in terms of the cosmological
constant\footnote{If $K_{\Lambda}:=\tilde{K}_n\Lambda^n$ represents
this expansion, it is found that
$\tilde{K}_3=K_3-{\textstyle{1\over3}}(*\,\varphi_G)^3$.}, and $J_3$
corresponds to the analytic part of the third coefficient of a similar
expansion of the Jones polynomial \cite{DiGr}. The identification of the
members of the family is then accomplished. Notice that the family is
composed in essence by those invariants involved (at third order) in the
relationship

\begin{equation}
K_q(\gamma) = q^{\textstyle{3\over4}\,\varphi_G}\,J_q(\gamma)
\label{kajones}
\end{equation}
between the Kauffman bracket and Jones knot polynomials in the variable
$q=e^{\Lambda}$ ($\Lambda$ is the cosmological constant and in
(\ref{kajones}) the vertical framing is assumed).

\subsection{The Mandelstam identities}
As it is known, the Kauffman bracket is formally given by the
expectation value of a Wilson loop in Chern-Simons theory
\cite{GaPubook}:

\begin{equation}
K_{\Lambda} = < W (\gamma) >_{\Lambda}
\label{kwil}
\end{equation}
This basic relation provides a natural explanation of the fact that the
knot invariants involved in the expansion of the Kauffman bracket in
terms of the cosmological constant satisfy the Mandelstam identities in
general. Under the light of (\ref{kwil}), this property is inherited
directly from the Wilson loops. In general, the verification of the
Mandelstam identities for diffeomorphism invariant expressions of the
general form (\ref{gwf}) is quite nontrivial. In the context of quantum
general relativity, the Mandelstam identities are basic requisites for
the wavefunctions in the loop representation.

The Mandelstam identities for the members of a family can be checked
systematically by means of equations (\ref{me1})-(\ref{me3}). However,
in the case of the family $\{\psi\}^6_4$ an immediate analysis follows
from the result (\ref{family}). In first place, all the members of a
family are cyclic by construction. For the first two families
$\{\psi\}^2_2$ and $\{\psi\}^4_3$, it is straightforward to see that the
invariants $J_2$ and $\varphi_G$ satisfy the requirements (\ref{me2})
and (\ref{me3}). The $*$-product of an arbitrary number of Gauss
invariants can be written in the following way:
$(*\,\varphi_G)^n=\varphi_G^n{_{\ldots}}\, X^{S[\cdots]}$, where
$S[\cdots]$ is the average over the totally symmetric permutation of
indices. This means that $(*\,\varphi_G)^n$ automatically satisfies the
symmetry identities (\ref{me2}) and (\ref{me3}) for any $n$. On the
other side, we know that
$K_3=\tilde{K}_3+{\textstyle{1\over3}}(*\,\varphi_G)^3$ with
$\tilde{K}_3$ the third coefficient of the expansion $K_{\Lambda}$ of
the Kauffman bracket in the cosmological constant. From the general
result (\ref{kwil}) we conclude that the analytic expression $K_3$ would
satisfy all the Mandelstam identities by construction. With respect to
$J_3$, the result (\ref{k3j3}) shows that the behaviour of this
invariant with respect to the second and third Mandelstam identities,

\begin{eqnarray}
(J_3)\,{_{\b \mu}}&\stackrel{?}{=}&
(J_3)\,{_{\bb \mu}}\label{jme2}\\
(J_3)\,{_{\b \alpha \b \beta \b \pi}}+
(J_3)\,{_{\b \alpha \b \beta \bb \pi}}
&\stackrel{?}{=}&(J_3)\,{_{\b \beta \b \alpha \b \pi}}
+(J_3)\,{_{\b \beta \b \alpha \bb \pi}}\,,
\label{jme3}
\end{eqnarray}
essentially depends of the properties of the product $J_2 *\varphi_G$
with respect to the same symmetry requirements. In general, the
overline symmetry (\ref{jme2}) is preserved under the $*$-product but
the identity (\ref{jme3}) is not. This means that $J_3$ would inherit
the property (\ref{jme2}) but not (\ref{jme3})\footnote{For ordinary
loops $J_3$ recovers the property (\ref{jme3}) for some privileged
topologies of the loop at the intersecting points. See \cite{Gr5} for a
more detailed discussion.}. The above discussion permits to conclude
that:

\begin{enumerate}
\item All the wavefunctions $\psi_1,\cdots,\psi_4$ given by expressions
(\ref{1})-(\ref{4}) are invariant under the overline operation
(\ref{jme2}).

\item Only $\psi_1$ and $\psi_2$ are invariant under the property
(\ref{jme3}).

\end{enumerate}
It is important to remark that for more complicated cases (i.e. for
higher ranks families) the analysis of the Mandelstam identities could
be worked in a systematic way.

\section{Conclusions and comments}
A systematic procedure to build up structured sets of diffeomorphism
invariants has been developed. The method is based on the properties of
the extended loop calculus.

A first point to remark is that the procedure of construction is {\it
generic} in the sense that it does not respond to any single topological
quantum field theory. In this respect, observe that the Chern-Simons
propagators can be considered merely a support for the extended loop
calculus and that a more basic mechanism (involving more general
objects) could underlie the analysis. Indeed, it could be possible that
a similar situation would occur with other building blocks (if exist)
for the wavefunction's propagators.

The fact that there is not a unique topological quantum field theory of
the Chern-Simons type underlying the procedure of construction has the
following important consequence: {\it the extended knot families would
contain all the diffeomorphism invariant expressions characterized by
the Chern-Simons propagators}. This fact opens new possibilities to
explore the analytic properties of knots. For the quantum gravity case
we have the additional benefit to dispose of a systematic control of the
Mandelstam symmetry requirements. The possibility of a systematic
analysis of the Mandelstam's identities for diffeomorphism invariants is
of relevance for the study of the Wheeler-DeWitt equation. Of particular
interest is the possibility to generate at higher rank families
invariants with the same analytic properties that the second coefficient
$J_2$. The $J_2$ invariant is the only known (nontrivial) analytic
expression that fulfills all the Mandelstam symmetry requirements and
that is annihilated by the vacuum Hamiltonian constraint (the
Br\"ugmann-Gambini-Pullin solution). Also, it is a true diffeomorphism
(ambient isotopic ) invariant not affected by framing ambiguities. In
the context of the extended knot families, $J_2$ is the first analytic
expression that appears after the most elementary Gauss invariant.

A more basic question is the problem to relate the individual solutions
of the Hamiltonian constraint with some knot polynomial. As we have
mention, $J_2$ is related to both Alexander-Conway and Jones
polynomials. The possibility that the expansion $J_{q}(\gamma)$ in the
variable $q=e^{\Lambda}$ could be the general solution of the
Wheeler-DeWitt equation without cosmological constant was suggested by
Br\"ugmann, Gambini and Pullin \cite{BrGaPu3}. Their conjecture was
mainly based on the formal relationship (\ref{kajones}) existing between
the Jones and the Kauffman bracket expansions. The analysis of this
conjecture to third order in the cosmological constant has shown that
the third coefficient of the Jones polynomial is not annihilated by the
vacuum Hamiltonian constraint in the general case \cite{Gr5}. On the
other hand, the Alexander-Conway knot polynomial is associated with the
expectation value of a Wilson type functional of a BF theory \cite{BF2}.
One can see that the topological quantity
$<W(\gamma)>_{\mbox{\scriptsize{BF}}}$ would satisfy a different sort of
Mandelstam identities that the usual Wilson loop expectation value. For
this reason, the knot invariants included in the perturbative expansion
of $<W(\gamma)>_{\mbox{\scriptsize{BF}}}$ could not be considered in
principle good candidates for quantum states of gravity. The above
punctuations stress again the role of the $J_2$ invariant. Is this
peculiar invariant a ``strange" (unique) case or there exist ``partners"
that share its same properties? The problem of the existence of a knot
polynomial associated with the space of states of quantum gravity
remains open. It is possible that a systematic study of extended knot
families could bring new insights to this question.

\section*{Acknowledgments}
I want to thank R. Gambini and J. Pullin for many fruitful discussions.
In special, I thank C. Di Bartolo by his comments and criticisms.

\section*{Appendix}
In this appendix we shall include a list of useful results. We have:

\begin{eqnarray}
&&\hspace{-1cm}
{\cal C}_{ax}\,C^1{_{\mu_1\ldots\mu_5}}X^{\mu_1\ldots\mu_5}=
\{ g_{\mu_4[\,ax}g_{bx\,]\,\mu_3}g_{\mu_1\mu_2}
+ g_{\mu_4[\,ax}g_{bx\,]\,\mu_1}g_{\mu_2\mu_3}
\nonumber\\
&&\hspace{-1cm}+g_{\mu_2[\,ax}g_{bx\,]\,\mu_1}g_{\mu_3\mu_4}
+\epsilon_{abc}(\phi^{\,cx}_{\,x_1}-\phi^{\,cx}_{\,x_4})
g_{\mu_1\mu_4}g_{\mu_2\mu_3}
+\epsilon_{abc}(\phi^{\,cx}_{\,x_1}-\phi^{\,cx}_{\,x_2}+\phi^{\,cx}_{\,x_3}
\nonumber\\
&&\hspace{-1cm}-\phi^{\,cx}_{\,x_4})g_{\mu_1\mu_2}g_{\mu_3\mu_4}
\} X^{(bx\,\mu_1 \mu_2 \mu_3\mu_4)_c}
+\{2g_{\mu_1[\,ax}h_{bx\,]\,\mu_2\mu_3}
\nonumber\\
&&\hspace{-1cm}+\epsilon_{abc}(\phi^{\,cx}_{\,x_3}-\phi^{\,cx}_{\,x_1})
h_{\mu_1\mu_2\mu_3}
+2\epsilon_{abc}\phi^{\,cx}_{\,z}
(\phi^{\,dz}_{\,x_3}-\phi^{\,dz}_{\,x_2})g_{\mu_1 \,dz}g_{\mu_2\mu_3}
\} X^{(bx\,\mu_1 \mu_2 \mu_3)_c}
\end{eqnarray}

\begin{eqnarray}
&&\hspace{-1cm}
{\cal C}_{ax}\,C^2{_{\mu_1\ldots\mu_5}}X^{\mu_1\ldots\mu_5}=
\{-g_{\mu_1[\,ax}g_{bx\,]\,\mu_3}g_{\mu_2\mu_4}
- g_{\mu_2[\,ax}g_{bx\,]\,\mu_3}g_{\mu_1\mu_4}
\nonumber\\
&&\hspace{-1cm}-g_{\mu_2[\,ax}g_{bx\,]\,\mu_4}g_{\mu_1\mu_3}
+\epsilon_{abc}(\phi^{\,cx}_{\,x_2}-\phi^{\,cx}_{\,x_3})
g_{\mu_1\mu_4}g_{\mu_2\mu_3}
+\epsilon_{abc}(\phi^{\,cx}_{\,x_1}+\phi^{\,cx}_{\,x_2}-\phi^{\,cx}_{\,x_3}
\nonumber\\
&&\hspace{-1cm}-\phi^{\,cx}_{\,x_4})g_{\mu_1\mu_3}g_{\mu_2\mu_4}
\} X^{(bx\,\mu_1 \mu_2 \mu_3\mu_4)_c}
-\{2g_{\mu_1[\,ax}h_{bx\,]\,\mu_2\mu_3}
\nonumber\\
&&\hspace{-1cm}+\epsilon_{abc}(\phi^{\,cx}_{\,x_3}-\phi^{\,cx}_{\,x_1})
h_{\mu_1\mu_2\mu_3}
+2\epsilon_{abc}\phi^{\,cx}_{\,z}
(\phi^{\,dz}_{\,x_3}-\phi^{\,dz}_{\,x_2})g_{\mu_1 \,dz}g_{\mu_2\mu_3}
\} X^{(bx\,\mu_1 \mu_2 \mu_3)_c}
\end{eqnarray}

\begin{eqnarray}
{\cal C}_{ax}\,C^1{_{\mu_1\ldots\mu_6}}X^{\mu_1\ldots\mu_6}&=&
\{ g_{\mu_2[\,ax}g_{bx\,]\,\mu_3}g_{\mu_1\mu_4}
\nonumber\\
&&+\epsilon_{abc}(\phi^{\,cx}_{\,x_3}
-\phi^{\,cx}_{\,x_2})
g_{\mu_1\mu_3}g_{\mu_2\mu_4}\}
X^{(bx\,\mu_1 \mu_2 \mu_3\mu_4)_c}
\end{eqnarray}

\begin{eqnarray}
{\cal C}_{ax}\,C^2{_{\mu_1\ldots\mu_6}}X^{\mu_1\ldots\mu_6}&=&
\{ g_{\mu_4[\,ax}g_{bx\,]\,\mu_1}g_{\mu_2\mu_3}
\nonumber\\
&&+\epsilon_{abc}(\phi^{\,cx}_{\,x_1}
-\phi^{\,cx}_{\,x_4})
g_{\mu_1\mu_2}g_{\mu_3\mu_4}\}
X^{(bx\,\mu_1 \mu_2 \mu_3\mu_4)_c}
\end{eqnarray}

\begin{eqnarray}
{\cal C}_{ax}\,C^3{_{\mu_1\ldots\mu_6}}X^{\mu_1\ldots\mu_6}&=&
\{ g_{\mu_2[\,ax}g_{bx\,]\,\mu_1}g_{\mu_3\mu_4}
+ g_{\mu_4[\,ax}g_{bx\,]\,\mu_3}g_{\mu_1\mu_2}
\nonumber\\
&&+\epsilon_{abc}(\phi^{\,cx}_{\,x_3}-\phi^{\,cx}_{\,x_2})
g_{\mu_1\mu_2}g_{\mu_3\mu_4}
\nonumber\\
&&+\epsilon_{abc}(\phi^{\,cx}_{\,x_1}
-\phi^{\,cx}_{\,x_4})
g_{\mu_1\mu_4}g_{\mu_2\mu_3}\}
X^{(bx\,\mu_1 \mu_2 \mu_3\mu_4)_c}
\end{eqnarray}

\begin{eqnarray}
&&{\cal C}_{ax}\,C^4{_{\mu_1\ldots\mu_6}}
X^{\mu_1\ldots\mu_6}=
\{ g_{\mu_1[\,ax}g_{bx\,]\,\mu_3}g_{\mu_2\mu_4}
+ g_{\mu_3[\,ax}g_{bx\,]\,\mu_2}g_{\mu_1\mu_4}
\nonumber\\
&&+ g_{\mu_2[\,ax}g_{bx\,]\,\mu_4}g_{\mu_1\mu_3}
+\epsilon_{abc}(\phi^{\,cx}_{\,x_3}-\phi^{\,cx}_{\,x_2})
g_{\mu_1\mu_4}g_{\mu_2\mu_3}
\nonumber\\
&&+\epsilon_{abc}(\phi^{\,cx}_{\,x_2}
-\phi^{\,cx}_{\,x_1}+\phi^{\,cx}_{\,x_4}
-\phi^{\,cx}_{\,x_3})g_{\mu_1\mu_3}g_{\mu_2\mu_4}\}
X^{(bx\,\mu_1 \mu_2 \mu_3\mu_4)_c}
\end{eqnarray}

\begin{eqnarray}
&&{\cal C}_{ax}\,C^5{_{\mu_1\ldots\mu_6}}
X^{\mu_1\ldots\mu_6}=
\{ g_{\mu_1[\,ax}g_{bx\,]\,\mu_4}g_{\mu_2\mu_3}
+ g_{\mu_1[\,ax}g_{bx\,]\,\mu_2}g_{\mu_3\mu_4}
\nonumber\\
&&+ g_{\mu_3[\,ax}g_{bx\,]\,\mu_1}g_{\mu_2\mu_4}
+g_{\mu_3[\,ax}g_{bx\,]\,\mu_4}g_{\mu_1\mu_2}
+ g_{\mu_4[\,ax}g_{bx\,]\,\mu_2}g_{\mu_1\mu_3}
\nonumber\\
&&+\epsilon_{abc}(\phi^{\,cx}_{\,x_2}-\phi^{\,cx}_{\,x_1}+\phi^{\,cx}_{\,x_4}
-\phi^{\,cx}_{\,x_3})(g_{\mu_1\mu_2}g_{\mu_3\mu_4}+
g_{\mu_1\mu_4}g_{\mu_2\mu_3})
\nonumber\\
&&+\epsilon_{abc}(\phi^{\,cx}_{\,x_1}-\phi^{\,cx}_{\,x_4})
g_{\mu_1\mu_3}g_{\mu_2\mu_4}\} X^{(bx\,\mu_1 \mu_2 \mu_3\mu_4)_c}
\end{eqnarray}

\end{document}